\documentclass[twocolumn,twoside,slac_two]{revtex4}
\usepackage{graphicx}
\usepackage{fancyhdr}
\begin{document}

\title{VERITAS Observations of Mkn 501 in 2009}

%

\author{Dongqing Huang, Alexander Konopelko}
\affiliation{Department of Physics, Pittsburg State University, 1701 South Broadway, Pittsburg, KS 66762}
\author{for the VERITAS collaboration}

\begin{abstract}
VERITAS is the high-sensitivity instrument of latest generation. It is often used for the short AGN monitoring exposures evenly distributed over entire observational season of a source of interest. Each of these exposures is long enough to detect the source 
at the flux level of about 1 Crab. During the 2009 observing season a number of exposures of Mkn~501 with VERITAS revealed variable TeV $\gamma$-ray emission at the flux level eventually exceeding 2 Crab. The spectral and flux variability measurements in TeV $\gamma$ rays for the 2009 data sample of Mkn~501 are summarized in this paper.
\end{abstract}

\maketitle

\thispagestyle{fancy}


\section{INTRODUCTION}

\noindent
Blazars are the most extreme variety of active galactic nuclei (AGN) known and remain one of the most intriguing classes of astrophysical objects. Their primarily non-thermal emission is characterized by irregular, rapid variability, core-dominated radio morphology, apparent superluminal motion, flat radio spectra and a broad continuum extending all the way from radio through $\gamma$ rays. Blazar properties imply relativistic beaming, i.e., the bulk relativistic motion of emitting plasma at small angles to the line of sight, which boosts a strong amplification and collimation in the observerÕs frame. Blazars are sites of high-energy phenomena, with a bulk Lorentz factor up to 50 and photon energies reaching the TeV energy range.

The BL Lac object Mkn~501 was first detected by the Whipple group~\cite{Quinn1996}. Early detections of Mkn~501 revealed a very low flux of TeV $\gamma$ rays, at the level of about 0.5 Crab. However, in 1997 Mkn~501 exhibited an unprecedented flare in TeV $\gamma$ rays with an integral flux of up to 10 Crab. A very long exposure on this source, lasting almost 6 months, yielded high statistics for TeV $\gamma$ rays, which provided very accurate measurements of the spectrum~\cite{Aharonian2001}. The spectrum of Mkn~501 was evidently curved and it was empirically fit by a power-law with an exponential cut-off. Occasionally Mkn~501 shows a very strong flux of X-ray emission. During the 1997 TeV flare the {\it Beppo}SAX satellite detected a dramatic increase in the X-ray flux up to 100 keV~\cite{Pian98}. Mkn~501 is a highly variable source of TeV $\gamma$-ray emission. The shortest flux variability discovered has a doubling time of a few minutes~\cite{Albert2007} . Such fast variability of the source is associated with sporadic changes of the flux level on much longer time scales. 

Mkn~501 has been the target of many multiwavelength campaigns mainly covering the object during flaring activity (see \cite{Kranich2009}). Simultaneous broadband observations of blazars in a flaring state provide an excellent test of emission models. Observations of blazars at GeV and TeV energies can profile the spectral shape of the high-energy component of their emission. As discussed in~\cite{K03}, the IR de-absorbed spectrum of BL Lac objects is rather flat at TeV energies and a high Lorentz factor $\gtrsim$50 is required to fit the observations in the inverse-Compton (IC) scenario \cite{K03}. Such an unusually high Lorentz factor, also constrained by recently established very short variability of the TeV $\gamma$-ray fluxes, is directly related to the formation and overall power of the AGN jet.

Here we report on recent observations of Mkn~501 taken with VERITAS in 2009 as a part of the large-scale multiwavelength campaign, joined by a number of ground-based and space-born experiments~\cite{Paneq09}.       

\section{VERITAS}
\begin{table*}[t]
{Table I Summary of Data}\\

\vspace*{1mm}

\begin{tabular}{|l|c|c|c|c|}
\hline \textbf{Setup} & 
\textbf{Period} & 
\textbf{Exposure [min]} &
\textbf{$\gamma$-Ray rate [$min^{-1}]$} &
\textbf{Significance [$\sigma$]} 
\\
\hline 4 telescopes & March 17 - 25, 2009 & 184 & 2.17$\pm$0.17 & 14.3\\
\hline 2 telescopes & April 30 - May 1, 2009 & 146 & 2.89$\pm$0.18 & 23.0\\
\hline 3 telescopes & May 29 - June 22, 2009 & 166 & 2.09$\pm$0.17 & 14.4\\
\hline
\end{tabular}
\label{table1}
\end{table*}

\noindent
VERITAS {\it (Very Energetic Radiation Imaging Telescope Array System)} is an array of four imaging atmospheric Cherenkov telescopes (IACT) located in southern Arizona at an altitude of 1.3~km~\cite{Holder06}. The telescopes are almost identical in their technical parameters. The 12~m optical reflector is a tessellated structure consisting of 357 identical spherical mirror facets, which are hexagonal in shape. The arrangement of the mirror facets constitutes a Davies-Cotton design, providing a total reflecting area of 110 $\rm m^2$. The optical point-spread function of a VERITAS telescope has a FWHM of about 4$^\prime$ on-axis. A high-resolution imaging camera placed at the focus of the reflector consists of 499 photomultiplier tubes (PMTs) in a close-packed hexagonal arrangement and has a field of view of $3.5^\circ$. Each camera PMT views a circle of diameter $0.15^\circ$ on the sky. A set of light concentrators is mounted in front of the PMTs to increase the light-collection efficiency and block the off-axis light. The camera triggers if the signal in each of any three adjacent PMTs exceeds a discriminator threshold of 50~mV, corresponding to approximately 4-5 photoelectrons. A coincidence of at least two cameras triggering within a time gate of 100~ns is required to read out an event.  The nominal trigger rate of the four-telescope array was about 230~Hz at zenith.

VERITAS is used to observe astrophysical sources from the northern hemisphere over the energy range from 100 GeV to 50 TeV, with a sensitivity of 7 mCrab (a $\gamma$-ray source of 0.7\% of the Crab Nebula flux will be detected with 5$\sigma$ significance over a 50 hour exposure). The measured Crab Nebula rate is 7 $\gamma$'s min$^{-1}$, which results in the significance of 31 $\rm \sigma\, hr^{-1/2}$. This detector could see a blazar flare at the 2 Crab level after a few minutes of observations. A unique sensitivity of the VERITAS instrument allows the investigation of the pattern of flux variability within the lifetime of a blazar flare in great detail. 
In addition, the outstanding capabilities of VERITAS as a stereoscopic array of IACT offer high-quality spectral measurements. However, due to a rather low duty cycle (10-12\%) and the narrow field of view of VERITAS, total available observing time is a subject of many, very competitive scientific projects. Therefore, scheduling of observing time for a variable source ultimately requires very careful consideration.

\section{OBSERVATIONS}

\noindent
Mkn~501 was observed with VERITAS for 8.3 hours between March 17 and June 22 of 2009. The observations were made with the full four-telescope array, as well as with the sub-arrays of three and two telescopes. During the nights of April 30 and May 1, two telescopes were disabled due to hardware issues leaving two operational telescopes from the full four telescope array. For the data in late May and during June, three telescopes were operational due to the move of telescope 1~\cite{Perkins09}. All data were taken in 20-minute runs using the standard {\it wobble} source-tracking mode, which is optimal for observations of a point-like source. In this observational mode the source is positioned at a 0.5$^\circ$ offset from the center of the field of view of the camera during observations, which allows for both on-source observations and simultaneous estimation of the background contamination caused by charged cosmic rays. 
Prior to applying analysis cuts, data were selected for adequate image quality. Each accepted event was also required to contain at least two images passing these cuts. Short summary of the observational data is given in Table~\ref{table1}. 

\begin{figure}[htbp]
\centering
\includegraphics[width=75mm]{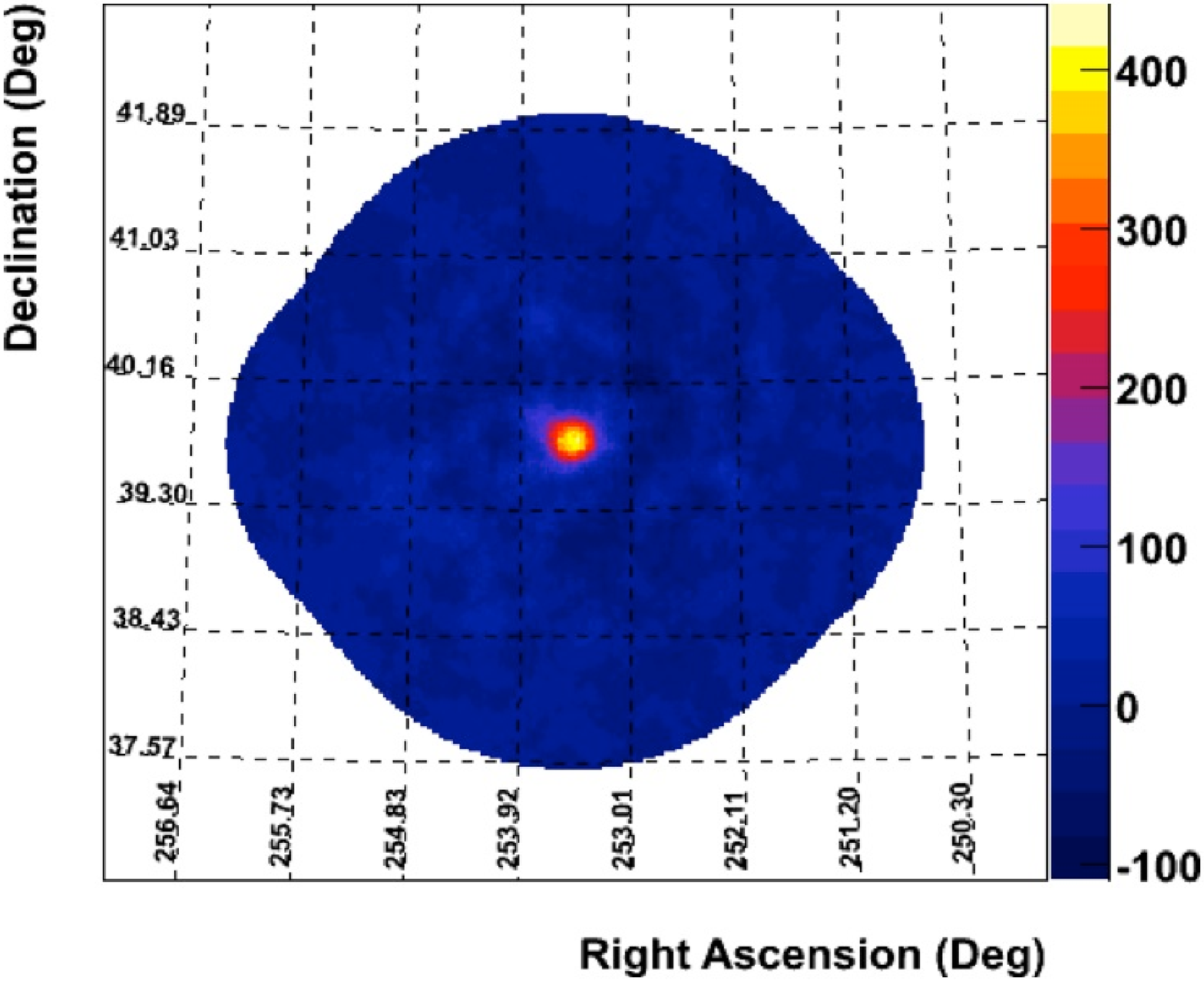}  

\vspace*{3mm}

\includegraphics[width=75mm]{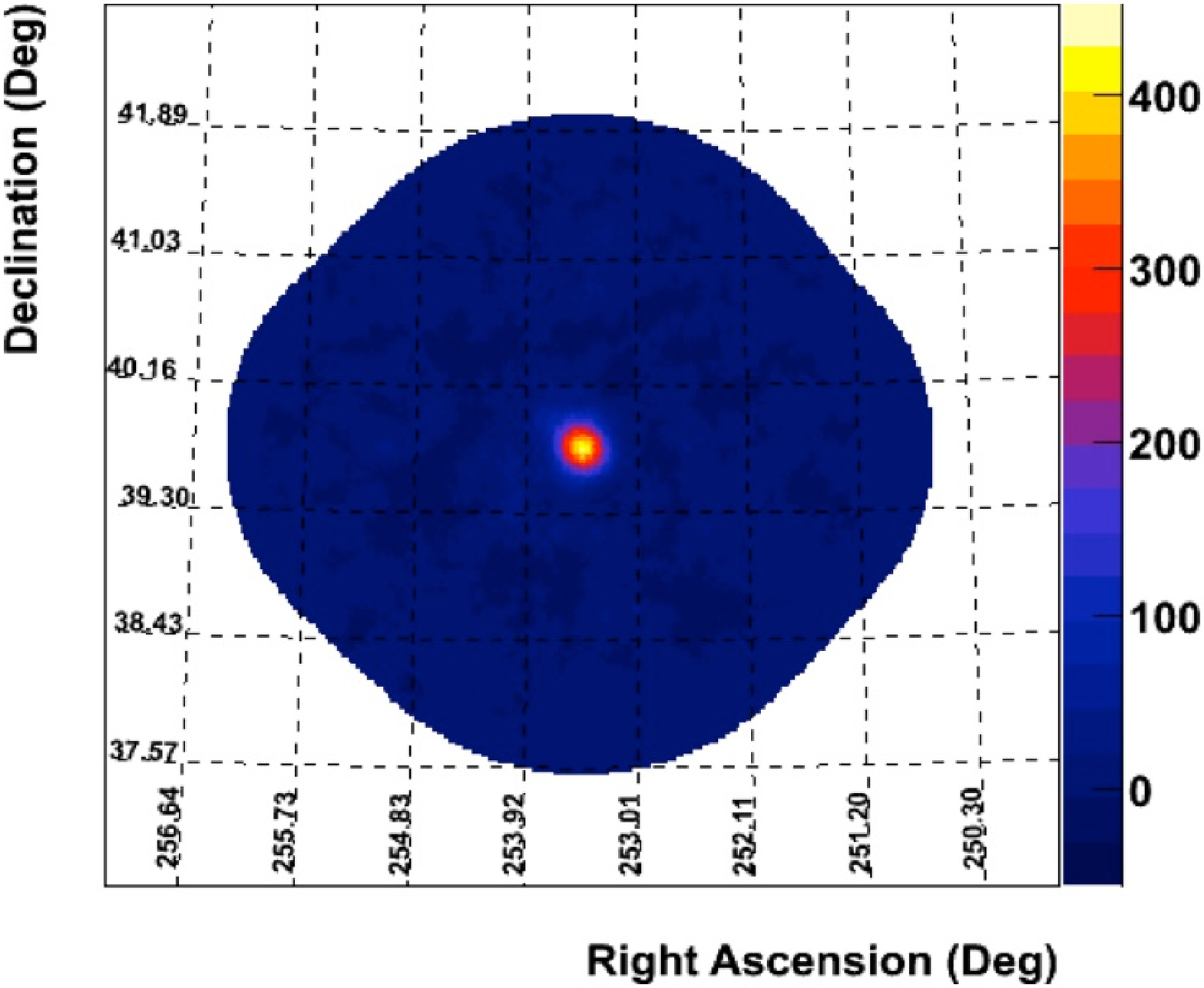}
\caption{Smoothed sky map of excess counts from the region centered at Mkn~501 
observed with VERITAS for 3 hours with 4 telescopes (upper panel) and for 2.4 hrs with 
2 telescopes (low panel). The color bar represents the excess event counts.} \label{f1}
\end{figure}

\section{ANALYSIS}

\begin{figure*}[t]
\centering
\includegraphics[width=120mm]{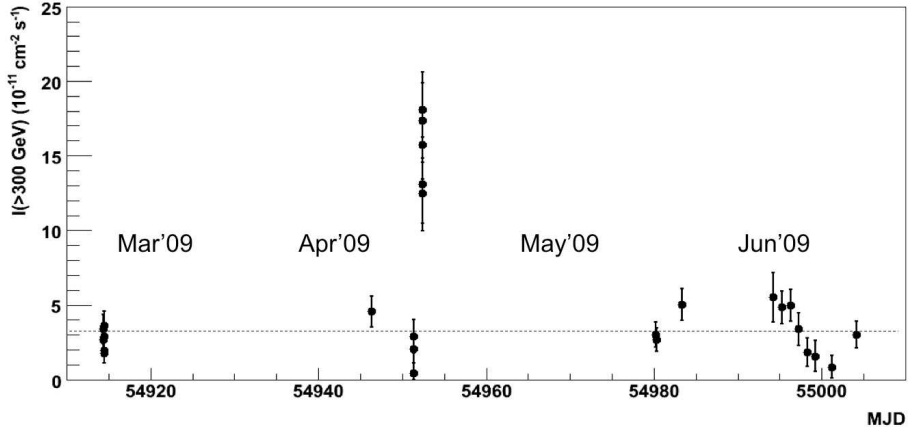} 
\caption{Run-by-run light curve of Mkn~501. The dashed line shows the fit to a constant integrated flux of the light curve excluding the nights of April 30 and May 1.} \label{f4}
\end{figure*}

\noindent
The imaging analysis of the VERITAS data is based on the reconstruction of the shower direction for each individual event and joint parametrization of the shape of the Cherenkov light flash from an individual shower using a multiple-telescope approach. All recorded events were subjected to the canonical directional cut $\theta^2$, where $\theta$ is the angular distance between the true source position on the sky and the reconstructed one. Of the remaining events, the candidates for $\gamma$-ray showers were selected using two simultaneously applied standard cuts on the parameters of image shape: MSW (mean scaled Width) and MSL (mean scaled Length). All recorded events that have passed both the primary image quality cuts and specific analysis cuts can be plotted in a two-dimensional sky map. The excess maps of the sky region around Mkn~501 for the data set of 3 hours taken during March 17 - 25, 2009 with four telescopes and for the data set of 2.4 hours taken during April 30 - May 1, 2009 with two telescopes are shown in Figure~\ref{f1}. An evident excess due to $\gamma$ rays at the high level of statistical significance can be observed at the position of Mkn~501 in both plots.

\section{RESULTS}

\noindent
Analysis of the Mkn~501 data, taken with the VERITAS array during 2009 observing season, revealed significant variations of the flux and noticeable change of the spectral slope at energies above 250 GeV depending on the emission state.  

\subsection{Light Curve}

\noindent
Figure~\ref{f4} shows the Mkn~501 light curve measured during March - June 2009. The integral $\gamma$-ray flux above 300~GeV was computed for each individual data run of 20~minutes. The light curve is consistent with the constant baseline emission of the source, excluding two days of observations in April 30 and May 1, when a prominent high-emission state (flare) was detected with VERITAS. The maximum $\gamma$-ray flux observed during the flare was $\rm I(>300\, GeV) = (1.80\pm0.25)\times 10^{-10} \rm [cm^{-2}s^{-1}]$, which is by a factor of 6 higher than the average baseline flux. Analysis of the intrarun time variability of the source in a flaring state is now ongoing.    

\subsection{Energy Spectrum}

\begin{figure}[t]
\centering
\includegraphics[width=75mm]{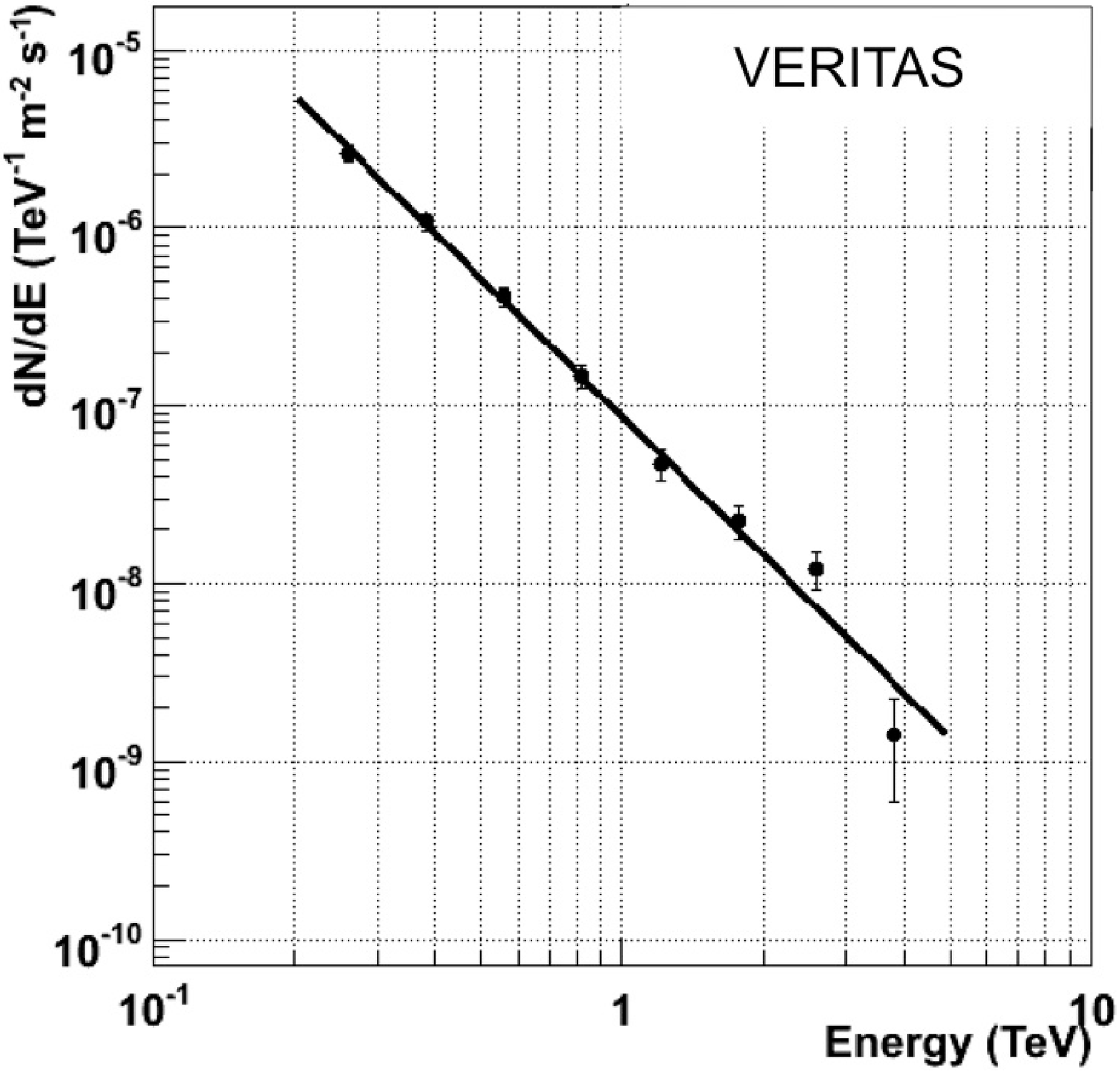} 

\vspace*{3mm}

\includegraphics[width=75mm]{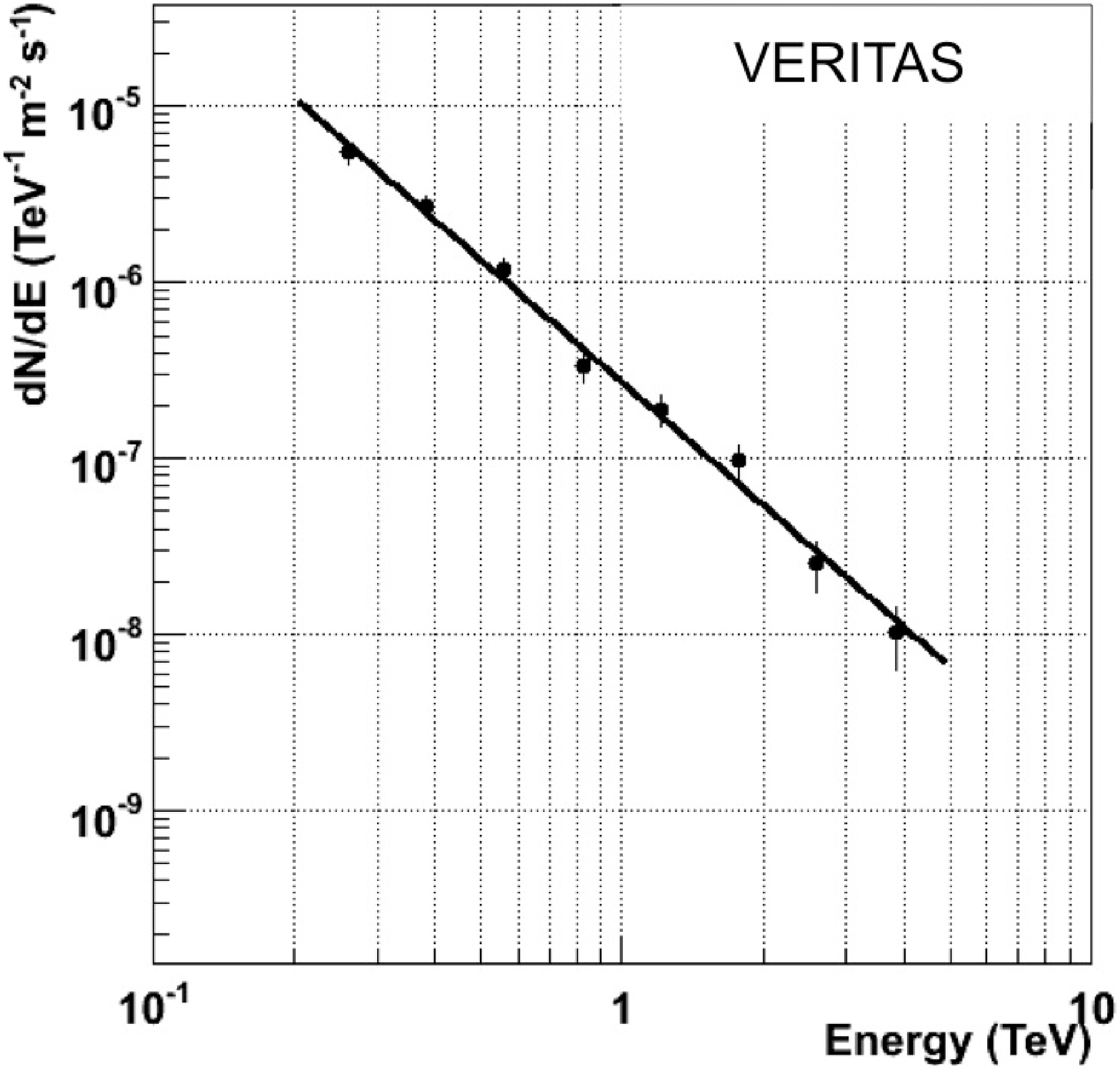}
\caption{Differential energy spectrum of TeV $\gamma$ rays from Mkn~501, measured with VERITAS. The upper panel shows the baseline emission spectrum evaluated using combined three and four telescope data samples. The low panel shows the high-flux spectrum for the two telescope data sample.
} \label{f2}
\end{figure}

\noindent
Evolution of the TeV energy spectrum across the short-time flaring state of Mkn~501 is of a great importance for understanding the mechanism of particle acceleration in a blazar-type source. The differential energy spectra of Mkn~501 ($dN_\gamma / dE$ $\rm [cm^{-2}s^{-1}TeV^{-1}]$) were evaluated for the baseline emission state 
\begin{equation}
\frac{dN_\gamma}{dE} =(8.7 \pm 0.6) \times 10^{-12} (E/1\, TeV)^{-2.58\pm 0.08 \pm 0.06} 
\end{equation}
as well as for the 2-days high emission state 
\begin{equation}
\frac{dN_\gamma}{dE} =(2.7 \pm 0.2) \times 10^{-11} (E/1\, TeV)^{-2.31\pm 0.08 \pm 0.06} 
\end{equation}
(also see Figure~\ref{f2}) during the 2009 observational campaign. Even though the spectrum indices noticeably differ, provided the statistical and systematic errors of the spectral indices, one can not claim any apparent spectral variability of Mkn~501 during this 2009 observational campaign. Further detailed analysis of this data set is required to draw a final conclusion on any substantial spectral change of TeV $\gamma$-ray emission from Mkn~501 in 2009. It is worth noting that the results reported in this paper are consistent with the previous observations of Mkn~501 at TeV energies made with VERITAS and the other ground-based experiments. 
 
The correlated analysis of the {\it Fermi} GST LAT and VERITAS data taken during 2009 observational season is currently ongoing and it will be presented elsewhere. Observations of blazar Mkn~501 at GeV and TeV energies can be used for profiling the spectral shape of the high-energy component of its emission. Simultaneous broadband observation of Mkn~501 in a flaring state provides excellent test of emission models. Multi-wavelength observations of Mkn~501 in the quiescent-state in March 2009 were summarized in~\cite{Gall09}.    

\section{SUMMARY}

\noindent
The first year of sky-survey operation with the {\it Fermi} GST LAT revealed a large number of bright blazars including many TeV BL Lacs. Several BL Lacs have been detected by ground-based TeV $\gamma$-ray instruments~\cite{Abdo09}. These objects are highly variable. High-energy $\gamma$-ray flares have been observed on time-scales of minutes to months. These flares vary greatly in amplitude. The high-energy $\gamma$-ray flares frequently correlate with variability seen at longer wavelengths, although the correlation is not always apparent. The origin of the blazar flares is still not well understood. They could be related to internal shocks in the jets, or they can be caused by a major ejection of a new component into a relativistic plasma flow of the jet. The high-energy component, seen by {\it Fermi} GST LAT and TeV ground-based detectors, is still a matter of considerable debate. 

All of the ground-based TeV $\gamma$-ray instruments continue further monitoring of well-established TeV blazars. This has paid off in a number of detected flares. The most recent flare of Mkn~501 was detected with VERITAS in April-May 2009. The flux of  TeV $\gamma$ rays has substantially exceeded the flux of the average baseline emission of Mkn~501 measured in 2009. The TeV $\gamma$-ray spectrum combined with the spectral measurements at other longer wavelengths can be used for detailed modeling of the broadband emission of the Mkn~501 blazar.     

\bigskip 
\begin{acknowledgments}

\noindent
This research was supported by grants from the U.S. Department of Energy, the U.S. National Science Foundation and the Smithsonian Institution, by NSERC in Canada, by Science Foundation Ireland and by STFC in the UK. The VERITAS collaboration acknowledges the hard work and dedication of the FLWO support staff in making the relocation of Telescope 1 possible. The VERITAS collaboration acknowledges the NASA support on the {\it Fermi} GST LAT Grant \#NNX08AV62G.
\end{acknowledgments}

\bigskip 

\end{document}